# New results on the Gunn-Peterson Effect at High Redshift

E. Giallongo[1], S. D'Odorico[2], A. Fontana[1], S. Savaglio[3], S. Cristiani[4], P. Molaro[5]

[1] Osservatorio Astronomico di Roma, I-00040 Monteporzio, Italy
[2] ESO, Karl Schwarzschild Str. 2, D-85748 Garching, Germany
[3] Dipartimento di Fisica, Università della Calabria, Italy
[4] Dipartimento di Astronomia, Università di Padova, I-35122 Padova, Italy
[5] Osservatorio Astronomico di Trieste, via G.B. Tiepolo 11, I-34131 Trieste, Italy

**Abstract.** Attempts to measure the optical depth due to a diffuse intergalactic medium (IGM) in the spectra of high redshift quasars (GP test) within the ESO key program on the intergalactic medium are reviewed. It is shown that there is no evidence for any Gunn-Peterson effect up to the highest redshifts observable in quasar spectra. The Ly$\alpha$ line statistics consistent with this limit is not able to explain the strong absorption observed at the HeII forest in one QSO, leaving room for a true HeII GP effect. The HI/HeII GP ratio implies a steep ionizing UV background at $z \sim 5$ and ionizing sources of stellar origin.

## 1 Introduction

The optical depth observed in spectra of high redshift quasars shortward of their Ly$\alpha$ emission is due to the presence of cosmologically distributed neutral hydrogen along the line-of-sight as suggested by Gunn & Peterson in 1965. The value they found was so low that it is generally assumed that the intergalactic medium is highly ionized rather than almost completely absent.

From an observational point of view, the Gunn-Peterson (GP) optical depth $\tau_{GP} = -ln(I_c/I_{extr})$ is the ratio between the local continuum level measured shortward of the quasar Ly$\alpha$ emission and the extrapolated continuum defined longward of the Ly$\alpha$ emission, where the quasar continuum emission is unaffected by intergalactic hydrogen absorption. In fact, most of the absorption is due to numerous and narrow absorption lines (the so called "Ly$\alpha$ forest") interpreted as HI Ly$\alpha$ absorption from intervening intergalactic clouds along the line-of-sight. Thus, the estimate of the true continuum level needed for the measure of $\tau_{GP}$ in the Ly$\alpha$ forest is made difficult by the large number of absorption lines present at high redshifts. In the next section, we briefly review previous measures on the GP effect that rely on the knowledge of the statistical properties of the Ly$\alpha$ lines as a function of redshift, and discuss the new results obtained by the authors within the ESO Key Project on the intergalactic medium at high z.

## 2  Estimates of the GP Optical Depth

Steidel & Sargent (1987) have shown that an upper limit to the diffuse HI absorption, relying on the knowledge of the average line absorption, can be obtained by subtracting the contribution of the Ly$\alpha$ lines to the total absorption observed between Ly$\alpha$ and Ly$\beta$ emissions in low-resolution QSO spectra. They estimated a value $\tau_{GP} \simeq 0.02 \pm 0.03$ at $z = 2.6$. Giallongo & Cristiani (1990) and Cristiani et al. (1993) compared all the available data with simulations of the average absorptions in synthetic QSO spectra computed on the basis of the known statistics of the Ly$\alpha$ lines. They showed that there was no evidence for a GP effect (i.e. $\tau_{GP} < 0.1$) up to $z = 5$. A more refined analysis, suggested by Jenkins & Ostriker (1991), consists in the relative frequency distribution of the transmitted fluxes. This way the simulated spectra, which depend on the line statistics, can be compared with a distribution of intensities rather than a single average value. In contrast, the observed distribution is influenced by the instrumental resolution. Thus, high resolution spectra are needed to derive a more sensitive test. Webb et al. (1992) applied a similar method to the high resolution spectrum of 0000-26 at $z_{em} = 4.1$. The inferred value was based on the slope $\beta$ of the column density distribution $\propto N_{HI}^{\beta}$ of the weak lines. They found $\tau_{GP} = 0.04$ for a steep $N_{HI}$ distribution with $\beta = -1.7$. A null optical depth required a flat $N_{HI}$ distribution with $\beta = -1.3$ extrapolated down to $\log N_{HI} = 12$. It is important to note in this respect that high resolution data at 14 km/s in the redshift interval $z = 2.9 - 3.6$ (Giallongo et al. 1993, Cristiani et al. 1995) indicate a flat $N_{HI}$ power-law with $\beta = -1.4, -1.5$ in the range $13.3 \leq \log N_{HI} \leq 14.5$ with a cutoff or a steepening at larger column densities. This favours the Webb et al. solution with $\tau_{GP} = 0$ at $z_{abs} = 3.8$.

A more stringent and direct upper limit to the GP effect has been given by Giallongo et al. 1992,1994 in the framework of an ESO key program devoted to the study of the intergalactic medium at high $z$. They used flux calibrated spectra obtained at relatively high resolution (14-40 km/s) and extending up to about 10000 Å (PKS 2126-158 $z_{em} = 3.3$; BR 1202-07 $z_{em} = 4.7$, respectively). The high resolution allows a better evaluation of the continuum shape and a direct selection of regions in the Ly$\alpha$ forest which are free of strong absorption lines. The ratio $I_c/I_{extr}$ in the Ly$\alpha$ forest was computed, giving an average optical depth at $z = 4.3$ $\tau_{GP} = 0.02 \pm 0.03$ where the error is due to the noise in the spectrum and to the slope uncertainty in the extrapolated continuum. The same method applied to the QSO PKS 2126-158 gives $\tau_{GP} \simeq 0.01 \pm 0.03$ at $z = 3$.

Using the Jenkins & Ostriker method, the histogram of the relative intensity in the Ly$\alpha$ forest of PKS 2126-158 in the range 4850–5000 Å is shown in Fig. 1a (thick line). The thin line is derived from a synthetic spectrum whose line distribution parameters are obtained from a maximum likelihood analysis on the high resolution line sample used by Cristiani et al. (1995). A best fit was obtained extrapolating the flat $N_{HI}$ power-law distribution with slope $\beta = -1.45$ down to $\log N_{HI} = 12.2$ with the addition of a GP optical depth $\tau_{GP} = 0.015$. The intensity histogram of the Ly$\alpha$ forest in BR 1202-07 is shown in Fig. 1b (thick line). The range is 6200–6450 Å, i.e. $z \sim 4.1 - 4.3$. The thin line is the extrapolation to $z = 4.3$ of the same $N_{HI}$ distribution obtained at $z = 3$. This extrapolation requires a steep power-law evolution in $z$ with slope $\gamma = 4.2$. Independent estimates by Zuo & Lu (1993) are consistent with increasing redshift evolution of the Ly$\alpha$ lines for $z > 4$. A best fit solution is found for low $N_{HI}$ thresholds in the range $\log N_{HI} = 12.6 - 12.8$ and with $\tau_{GP} = 0 - 0.04$. The GP optical depth found in this way is consistent within $1\sigma$ with the value measured in the Ly$\alpha$ forest by Giallongo et al. (1992,1994) selecting regions of width 3-4 Å which are free of strong absorption lines. Fig. 2a shows that intervals between Ly$\alpha$ lines of this width are not rare voids, but are just expected from a poissonian distribution of the same synthetic lines ($\log N_{HI} > 12.6$) which fit the observed histogram of the pixel intensity.

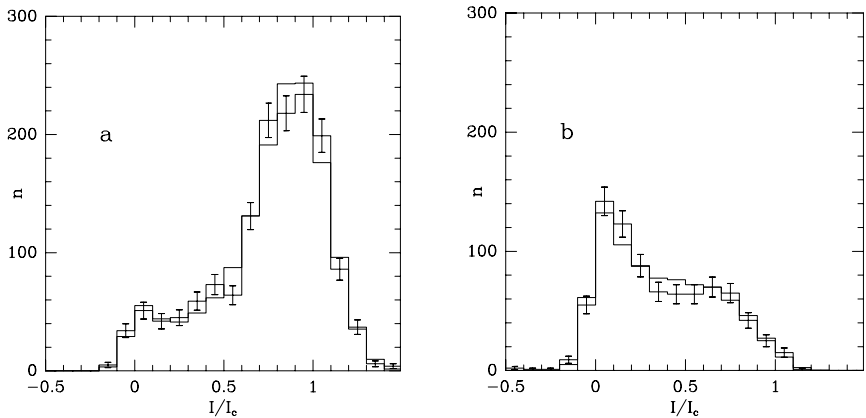

**Fig. 1.** Intensity histograms of the Ly$\alpha$ forest in PKS 2126-158 a), and BR 1202-07 b), (thick lines). Best fit intensity distributions from corresponding synthetic spectra are also shown (thin lines). In a) $\tau_{GP} = 0.015$; in b) $\tau_{GP} = 0$.

It is interesting to show the contribution of the same synthetic line distributions to the average optical depth at the HeII Lyman-$\alpha$ forest, assuming an intensity ratio $S_L$ of the UV background between the Lyman limit and the HeII edge $\sim 10^2$. The predicted intensity distribution at $z = 3.3$ is shown in Fig. 2b. The corresponding average $\tau_{HeII} = 1.2$ so obtained, can be compared with the strong absorption ($\tau_{HeII} = 3.2$) observed for one line-of-sight at the same $z$ by Jakobsen et al. (1994). Increasing $S_L$ to values $\sim 10^3$ does not change appreciably the average HeII optical depth ($\tau_{HeII} = 1.6$, (see also Madau & Meiksin 1995)). The two HI line distributions used to fit the data at $z = 4.1 - 4.3$ give a contribution in the HeII forest $\tau_{HeII} = 2.5 - 2.1$, (with $S_L = 10^2$), respectively. Thus, the Ly$\alpha$ line population can not reproduce the observed HeII optical depth and a true $\tau(HeII)_{GP} \sim 1.8$ optical depth is suggested (using the best fit value given by Jakobsen et al. 1994). This high value remains consistent with the small upper limit for the GP optical depth found in the HI Ly$\alpha$ forest if $S_L \sim 200$.

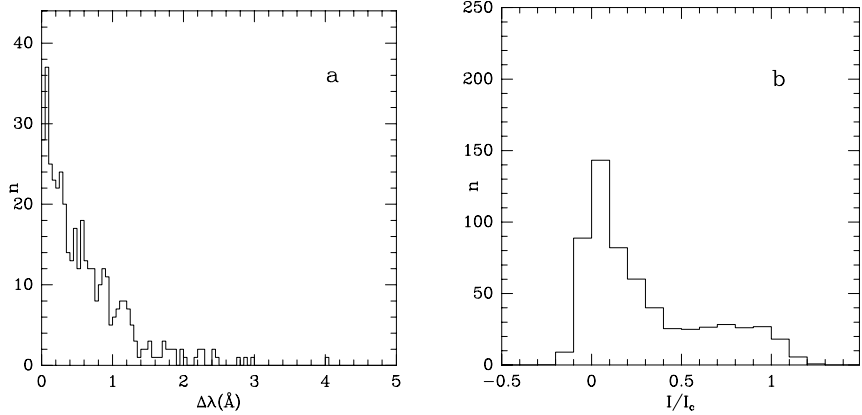

**Fig. 2.** a) Interval distribution in a synthetic spectrum (log $N_{HI} \geq 12.6$) which fits the intensity histogram in Fig. 1a. b) Intensity histogram in the HeII Ly$\alpha$ forest of the same synthetic spectra as in Fig. 1a.

A dominant starlight contribution to the UVB at $z \sim 5$ seems to be favoured both by the high value of the UVB at the Lyman limit (needed to explain the small value $\tau_{GP} < 0.04$ found at $z \sim 4.5$) and by the amount of steepening of the UVB between HI Lyman limit and the HeII edge (needed to explain the high value $\tau(HeII)_{GP} \sim 1.8$ at $z = 3.3$).